# Controlling the propagation of dipole-exchange spin waves using local inhomogeneity of the anisotropy

Morteza Mohseni*, Burkard Hillebrands, and Philipp Pirro

Fachbereich Physik and Landesforschungszentrum OPTIMAS, Technische Universität Kaiserslautern, 67663 Kaiserslautern, Germany

Mikhail Kostylev
Department of Physics and Astrophysics M013, University of Western Australia, 35 Stirling Hwy, 6009 Western Australia, Australia

Spin waves are promising candidates to carry, transport and process information. Controlling the propagation characteristics of spin waves in magnetic materials is an essential ingredient for designing spin-wave based computing architectures. Here, we study the influence of surface inhomogeneities on the spin wave signals transmitted through thin films. We use micromagnetic simulations to study the spin-wave dynamics in an in-plane magnetized yttrium iron garnet thin film with thickness in the nanometre range in the presence of surface defects in the form of locally introduced uniaxial anisotropies. These defects are used to demonstrate that the Backward Volume Magnetostatic Spin Waves (BVMSW) are more responsive to backscattering in comparison to Magnetostatic Surface Spin Waves (MSSWs). For this particular defect type, the reason for this behavior can be quantitatively related to the difference in the magnon band structures for the two types of spin waves. To demonstrate this, we develop a quasi-analytical theory for the scattering process. It shows an excellent agreement with the micromagnetic simulations, sheds light on the backscattering processes, and provides a new way to analyze the spin-wave transmission rates in the presence of surface inhomogeneities in sufficiently thin films, for which the role of exchange energy in the spin wave dynamics is significant. Our study paves the way to designing magnonic logic devices for data processing which rely on a designed control of the spin-wave transmission.

## I. Introduction

The field of magnonics is concerned with studying spin waves (SW), whose quanta are known as magnons. SWs are the low-energy eigen-excitations of magnetic materials [1]. The use of SWs promises many opportunities for data processing using wave-based, quantum- and unconventional computing concepts [2–5]. SWs are particularly interesting for several reasons: For instance, they provide access to nanometer-range wavelengths [6,7], feature a large range of operational frequencies from GHz to THz [8], and are also characterized by the absence of Ohmic-loss contributions to their dynamics, and hence the absence of Joule heating [1,9]. Therefore, using SWs opens up potential possibilities for the design and implementation of energy-efficient nanometer-scale device architectures for data processing that promise to become even more compatible with the CMOS platform in the years to come.

Controlling magnon transport is a necessary task for using SWs as data carriers. So far, several mechanisms have been used for this purpose. One can refer to, for instance, employing voltage [10–12] and strain [13,14], tuning of the magnon chemical potential via non-local injection [15,16], laser beams [17], spin textures [18–20], etc. Recently, it was shown that surface inhomogeneities in the form of surface roughness and defects can significantly influence the transmission characteristics of the SWs in thin films [21–24]. In those films, the contribution of the exchange energy to the magnon band structure is large enough to suppress direct mode-hybridization by opening up an energy gap between the fundamental mode and higher-order thickness modes [25]. In this context, SWs in such systems are very interesting for applications due to their dimensions and the facile control of their wave vector.

Here, we use a different type of surface inhomogeneity to tune the propagation characteristics of the SWs in an in-plane magnetized thin film. We utilize locally introduced uniaxial anisotropies at the surfaces of a thin film ("anisotropy defects") and demonstrate theoretically and numerically that transmission of a SW signal can be controlled by the defect strength. Unlike topographical defects [21], this type of inhomogeneity is analogous to an internal boundary which shifts the frequency of the local band structure of the magnons over the entire wave vector range in the defect region and thus requires a different approach to characterize the transmission rates of the SWs. Moreover, studying these types of surface inhomogeneities is important for several reasons: (*i*) magnetic inhomogeneities and material defects are unavoidable during thin film deposition. This can, for example, lead to linear magnon scattering processes [26]; (*ii*) it has been shown that breaking the spatial symmetry of the system with a certain periodicity by varying the effective field or the saturation magnetization, can lead to the presence of band gaps in the transmission characteristics of the SWs. This phenomenon is known as the magnonic crystal [27–29]; (*iii*) the capability to control the transmission characteristics of SWs using surface defects may represent an extra tool for designing magnonic devices that are essential for data processing using waves, such as phase shifters [30].

We use micromagnetic simulations and quasi-analytical calculations to study SW propagation in an yttrium iron garnet (YIG) thin film slab that is continuous in the slab plane and has a continuous thickness $t = 80$ nm. By introducing local uniaxial anisotropy at the surfaces of the film, we investigate how SWs are transmitted through and reflected from such defects. We define the local uniaxial anisotropies in a way that they effectively act as local variation in the effective field or the effective magnetization inside the inhomogeneities without changing the saturation magnetization. This leads to a local shift of the SW dispersion, and, consequently, affects the SW transmission rates. We show that based on the notion of a local dispersion relation inside the defect region, it is possible to explain the propagation and reflection mechanisms of the SWs. Furthermore, by comparing the Magnetostatic Surface Spin Waves (MSSWs) and Backward Volume Magnetostatic Spin Waves (BVMSW), we demonstrate how the difference in their magnon band structures impacts the SW propagation through the defects. We note that we use YIG as the modeled film material because YIG exhibits the lowest Gilbert damping known so far. Therefore, YIG is the most promising host material for SWs [31–33]. However, we note that the reported results can also be applied to metallic magnetic alloys such as NiFe and CoFeB.

## II. Methods

### A. Micromagnetic simulations

Micromagnetic simulations have been carried out using the MuMax 3.0 open source GPU-based package [34]. This package uses the finite difference method in order to solve the Landau-Lifshitz-Gilbert equation on a rectangular mesh,

$$\frac{d\vec{m}}{dt} = -\gamma \frac{1}{1+\alpha_{Gilbert}^2} (\vec{m} \times \vec{B}_{\text{eff}} + \alpha_{Gilbert}(\vec{m} \times (\vec{m} \times \vec{B}_{\text{eff}}))). \qquad (1)$$

Here $\vec{m}$ is the magnetization vector, $\gamma$ is the gyromagnetic ratio, $\alpha_{\text{Gilbert}}$ is the Gilbert damping constant, $\vec{B}_{\text{eff}}$ is the effective field including the external, exchange, magnetostatic and anisotropy fields. The system under investigation is a thin film slab with dimensions equal to 50 μm × 1 μm × 80 nm (length × width × thickness $t$) as shown in Fig. 1a. The system is divided into 5000 × 4 × 10 mesh cells leading to cell sizes of 10 nm × 250 nm × 8 nm. We use periodic boundary conditions (PBC×100) in the $y$-direction, which means that the film is practically infinite in the $y$-direction. Absorbing boundaries are assumed at the ends of the film in order to prevent back-reflections and interference effects, and realistic magnetic parameters for a YIG thin film are employed: $M_s = 140$ kA/m, $\alpha_{Gilbert} = 0.0002$, $A_{exch} = 3.5$ pJ/m [35]. We set the external field along the $y$-direction to simulate MSSWs, while for BVMSW, the external field is applied along the $x$-direction. The field amplitude is set to $\mu_0 H_{ex} = 0.05$ T in all simulations. A SW excitation source with a width of 20 nm (along the $x$-direction) and a height equal to the thickness (80 nm) is introduced at the middle of the film length ($x = 0$ μm) to generate SWs. A spatially homogeneous and sinusoidal oscillating field pulse with a frequency $f$ and an amplitude equal to 0.1 mT for $t = 10$ ns is applied perpendicular to the static field in the excitation source region. Once the excitation pulse is over, the generated SWs propagate for 60 ns to ensure that the entire wave packet reaches the two opposite edges of the slab. The collected simulation data is analyzed using the fast Fourier transform (FFT) method in space and time. To this end, dynamic magnetization components of the second layer along $y$ ($m_x$ for MSSWs and $m_y$ for BVMSWs) are collected through the entire simulation time.

We define the SW transmission as:

Transmission (%) = $\frac{[m_d(f,x)]}{[m_r(f,x)]} \times 100$ (2)

Here $[m_d(f, x)]$ and $[m_r(f, x)]$ refer to the FFT amplitude of the SWs at a location $x$ past the defect region for the film with the defect, and the FFT amplitude for a reference film without defect at the same location $x$, respectively.

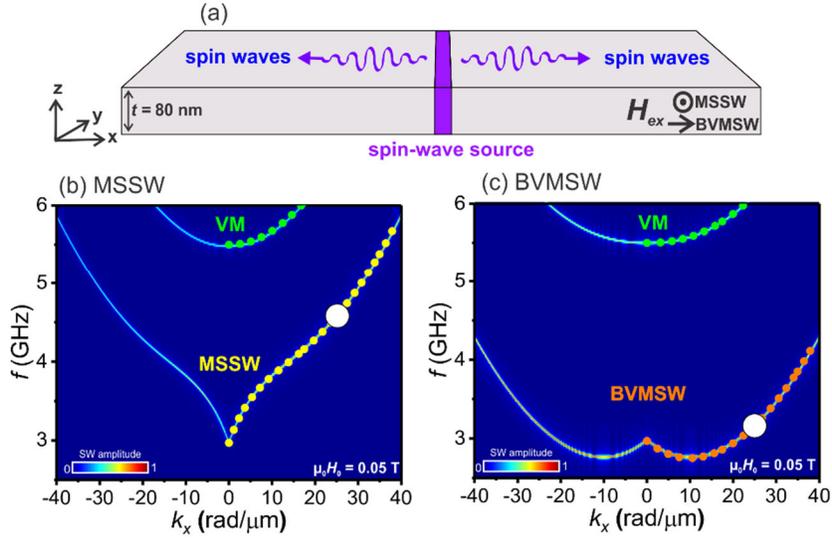

FIG 1. (a) Schematic picture of the systems under study, a YIG thin film with a thickness $t$. The film is magnetized in its plane in one of two possible directions with respect to the wave vector, corresponding to MSSW and BVMSW. (b-c) Magnon band structures for the MSSW and BVMSW modes (mode $n=0$ in Eq. (12)) for the reference film calculated via numerical simulations (colour plot) and analytically. The MSSW and BVMSW modes are shown with yellow and orange color, respectively; higher-order volume modes (VM, mode $n=1$ in Eq. (12)) are indicated by the green lines.

**B. Original Born-Equation based model of spin wave scattering from 2D defects**

Simultaneously to carrying out the micromagnetic simulations, we also developed a problem-oriented quasi-analytical theory of SW scattering from 2D defects. The theory assumes that both the film and the defect are continuous in the *y*-direction. We consider single-frequency plane SW with wave fronts of infinite length in the *y* direction.

As shown in Refs. [36–38], the problem of scattering of SW from 1D defects can be cast in the form of a Born Equation. The Born Equation is an integral equation for the amplitude of a wave that undergoes scattering from an area (called a "defect" in the present manuscript), where the value of some parameter of the waveguiding medium is different from its value elsewhere. In Ref. [36], the defect occupied the whole thickness of the film and the strength of the parameter (the strength of an additional localized external field in that paper) was almost uniform across the film thickness. Furthermore, the experiment carried out in the same paper was concerned with transmission of the fundamental thickness mode of SWs. For these two reasons, a rather accurate description of the problem was obtained with the simple method of assuming that the magnetization dynamics are thickness- (i.e. *z*-direction in Fig.1(a)) uniform and averaging all effective dynamic fields as well as the extra defect field over the film thickness. This yielded a one-dimensional integral equation of a very simple form.

An important advantage of the Born method is that the integral equation's area of definition of is the volume of the defect, and the areas outside the defect volume are excluded from consideration. One simple way to solve the integral equation is by using numerical methods. In Ref. [36] this was needed because the shape of the defect was complex. In these circumstances, the advantage of the integral-equation formulation of the scattering problem becomes very pronounced – the discrete mesh needs to cover the defect area only, and, thus a small number of mesh cells is sufficient to solve the problem numerically with good accuracy. Furthermore, the problem can be formulated in the frequency domain – by considering a plane single-frequency wave incident onto the defect. This results in a very efficient algorithm that delivers a solution within a minute. Conversely, the standard micromagnetic software, such as MuMax or OOMMF, needs a mesh that covers not only the defect areas, but also significant lengths in front and behind the defect. Furthermore, these packages are not problem-oriented and have a lot of built-in redundancies. This slows down the simulations. These unnecessary features are: (*i*) the necessity to simulate the static magnetization configuration first before starting simulating the SW dynamics; (*ii*) the SW nonlinearity; (*iii*) the intrinsic three-dimensionality of the problem formulation; and (*iv*) the necessity to solve the problem in the time domain for pulses of finite length with subsequent Fourier analysis of the simulated data. All these peculiarities slow down the simulations by a lot – a typical simulation run for our geometries takes approximately two hours.

Contrary to Ref. [36], the present paper considers defects that do not occupy the whole film thickness (in the direction *z*), therefore the present problem is essentially two-dimensional. Here we extend the theory to the second dimension. In order to obtain simple expressions that allow easy conversion into a short and fast numerical code, we use the fact that we deal with ultra-thin films.

Similar to the previous treatment [36], to formulate the scattering problem, we cast the linearized Landau-lifshitz equation in the following form

$$\hat{\chi}(\omega)^{-1}\mathbf{m}(x,z) = \mathbf{h}_d(x,z) + \mathbf{h}_{exc}(x,z) + \mathbf{h}_{ani}(x,z) + \mathbf{h}_{dr}(x,z), \quad (3)$$

where $\hat{\chi}(\omega)$ is the microwave magnetic susceptibility tensor [39] that we assume to be uniform over the whole volume of the film; $\mathbf{h}_d$ is the dynamic dipole field of precessing magnetization; $\mathbf{h}_{exc}$ is the dynamic effective field of (inhomogeneous) exchange interaction; $\mathbf{h}_{ani}$ is the effective anisotropy field; and $\mathbf{h}_{dr}$ is a spatially-localized microwave magnetic field of frequency $\omega$ that excites a SW incident onto the defect. The dynamic dipole field is obtained by solving Maxwell Equations in the magnetostatic approximation.

The solution takes a simple form in the Fourier space [40],

$$\mathbf{h}_{dk}(z) = \int_0^t \hat{G}_k(z-z')\mathbf{m}_k(z')\mathrm{d}z', \qquad (4)$$

where $t$ is the film thickness,

$$\mathbf{h}_{dk}(z) = \frac{1}{2\pi}\int_{-\infty}^{\infty} \mathbf{h}_d(x,z)\exp(ik_x x)\mathrm{d}x, \qquad (5)$$

$$\mathbf{m}_{dk}(z) = \frac{1}{2\pi}\int_{-\infty}^{\infty} \mathbf{m}(x,z)\exp(ik_x x)\mathrm{d}x, \qquad (6)$$

and $\hat{G}_k(z)$ is the Fourier-space Green's Function of the dipole field, whose components are shown in [40]. The Fourier image of the effective exchange field can be written down as [40],

$$\mathbf{h}_{exck}(z) = \alpha[\partial^2 \mathbf{m}_k(z)/\partial z^2 - k_x^2 \mathbf{m}_k(z)], \qquad (7)$$

where $\alpha$ is the inhomogeneous-exchange constant that we assume to be the same for any point within the film. We introduce the effective field of anisotropy employing the tensor of effective demagnetizing factors of anisotropy $\hat{N}_{ani}$ [41]. Because the tensor is local, this must be done in the real space

$$\mathbf{h}_{ani}(x,z) = \hat{N}_{ani}(x,z)\mathbf{m}(x,z). \qquad (8)$$

We start with solving the SW excitation and propagation problem for a homogeneous film, i.e. a film without defect. In this case, $\hat{N}_{ani} = \hat{N}_{ani}^0$ is co-ordinate independent and (8) reduces to

$$\mathbf{h}_{ani}(x,z) = \hat{N}_{ani}^0 \mathbf{m}(x,z). \qquad (9)$$

Substitution of (4-7) and (9) into (3) yields an integro-differential equation with respect to $\mathbf{m}(x,z)$. The differential part of the equation originates from the second-derivative term in (7) and the integral part from the Green's function formulation of the dipole field. The presence of the differential requires boundary conditions for the exchange-field operator (7) at the film surfaces $z=0$ and $z=t$. We assume the "unpinned surface magnetization" boundary conditions (see e.g. [40])

$$\partial \mathbf{m}(x,z)/\partial z \big|_{z=0,t} = 0. \qquad (10)$$

Under this assumption, we may cast the solution of the integro-differential equation into the form

$$\mathbf{m}(x,z) = \sum_{n=0}^{N} \mathbf{m}_n(x)\cos(n\pi z/t). \qquad (11)$$

As shown in [25], for ultra-thin films, it is sufficient to keep the first two terms ($n=0$ and $n=1$) of the expansion. This yields a solution in a very simple form

$$\mathbf{m}(x,z) = \mathbf{m}_0(x)\frac{1(z)}{\sqrt{t}} + \sqrt{\frac{2}{t}}\mathbf{m}_1(x)\cos(\pi z/t), \qquad (12)$$

where

$$1(z) = \begin{cases} 1, & 0 \le z \le t \\ 0, & \text{elsewhere} \end{cases}, \qquad (13)$$

and we introduced the pre-factor $\sqrt{2}$, in order to properly normalize our orthogonal basis functions $f_n(z)$ ($n=0,1$):

$$\int_0^t [f_0(z)]^2 \, dz = \int_0^t \left[\frac{1(z)}{\sqrt{t}}\right]^2 dz = 1$$

$$\int_0^t [f_1(z)]^2 \, dz = \int_0^t \left[\sqrt{\frac{2}{t}}\cos(\pi z/t)\right]^2 dz = 1 \qquad (14)$$

Substituting (12) into the integro-differential equation and projecting the resulting equation onto the orto-normal basis (14) yields a system of two linear integral equations for $\mathbf{m}_0(x)$ and $\mathbf{m}_1(x)$. Now we introduce a column-vector in a Hilbert space

$$|\mathbf{m}(x)\rangle = \begin{pmatrix} \mathbf{m}_0(x) \\ \mathbf{m}_1(x) \end{pmatrix}. \qquad (15)$$

This converts the system of the integral equations into an integral equation with a kernel in the form of a tensor

$$\hat{\chi}^{-1}|\mathbf{m}(x)\rangle - \int_{-\infty}^{\infty} \hat{C}(x-x')|\mathbf{m}(x')\rangle \, dx' = |\mathbf{h}_{dr}(x)\rangle, \qquad (16)$$

where the vector $|\mathbf{h}_{dr}(x)\rangle$ represents projections of the driving field onto the orthogonal basis (14), and the components of the tensor

$$\hat{C}(x) = \begin{pmatrix} C_{00}(x) & C_{01}(x) \\ C_{10}(x) & C_{11}(x) \end{pmatrix} \qquad (17)$$

are combinations of projections of the operators (4), (7) and (9) onto the orthogonal basis.

Equation (16) admits a formal solution

$$|\mathbf{m}(x)> = \int_{-\infty}^{\infty} \hat{G}_{exc}(x-x')|\mathbf{h}_{dr}(x)> dx', \qquad (18)$$

where $\hat{G}_{exc}(x)$ is a Green's function of SW excitation by an external driving field. Its derivation is fully analogous to the one-dimensional cases [36,37]. Equating the denominator of the components of the Fourier transform of $\hat{G}_{exc}(x)$ to zero yields dispersion relations $\omega(k_x)$ for the fundamental and the first exchange SW modes. (We will obtain the dispersion relations in a slightly different way in Section II B.)

Let us now introduce a defect in the form of a localized non-uniformity of the effective anisotropy field

$$\hat{N}_{ani}(x,z) = \hat{N}^0_{ani} + \delta\hat{N}_{ani}(x,z), \qquad (19)$$

where $\delta\hat{N}_{ani}(x,z) = \begin{cases} \delta\hat{N} & \text{inside the defect} \\ 0 & \text{elsewhere} \end{cases}$, (20)

and $\delta\hat{N}$ is a constant tensor that describes the defect amplitude.

With (19), Eq. (3) transforms into

$$\hat{\chi}^{-1}|\mathbf{m}(x)> - \int_{-\infty}^{\infty} \hat{C}(x-x')|\mathbf{m}(x')> dx' + \delta\hat{N}_{ani}|\mathbf{m}(x)> = |\mathbf{h}_{dr}(x)>. \qquad (21)$$

We now multiply both sides of (21) by $\hat{G}_{exc}(x''-x)$ and integrate over $x$ from the negative infinity to the positive one. With (18), this yields

$$|\mathbf{m}(x)> + \int_0^d \hat{G}_{exc}(x-x')\Delta\hat{N}|\mathbf{m}(x')> dx' = |\mathbf{m}^0(x)> \exp(-ik_x x)\exp(-\nu x). \qquad (22)$$

(Note that we replaced $x''$ with $x$ and $x$ with $x'$ in order to arrive at this final form.)

This is the sought Born Equation for scattering of SWs from 2D defects. In this equation, $\Delta\hat{N}$ is a tensor that is obtained by projecting $\delta\hat{N}_{ani}(x,z)$ onto the orthogonal basis (14), and

$$|\mathbf{m}^0(x)> \exp(-ik_x x)\exp(-\nu x) \qquad (23)$$

is the SW incident onto the defect, $|\mathbf{m}^0(x)>$ is its amplitude, $k_x$ is its wave number satisfying the dispersion relation $\omega(k_x)=\omega$, $\omega$ is the frequency of the driving field (see Eq.(3)), and $\nu$ is the decrement of spatial decay of SW. Note that $|\mathbf{m}^0(x)> \exp(-ik_x x)\exp(-\nu x)$ represents the exact solution of (18) for the far zone of the excitation source [36]. In addition, note that the decrement of the spatial decay of SW appears in this expression because we take into account magnetic losses while writing down the expression for $\hat{\chi}(\omega)$ [39]. For the same reason, the Green's function in Eq. (22) takes into account the SW decay. Thus, Eq.(22) accounts for the effect of SW decay on the efficiency of SW scattering from the defect.

The physical meaning of the equation is that every point of the defect *x'* represents a source of secondary waves. These waves combine (through the mathematical operation of integration) at any point *x* to produce a scattered SW field. The scattered field is given by the integral term of the equation (with the negative sign). The scattered field for locations *x*<0 in front of the defect (i.e. the back-scattered field) represents the reflected SW, and the combination of the incident wave (given by the r.h.s. term of the equation) and the forward-scattered field represents the transmitted wave for locations behind the defect (*x*>*d*).

Importantly, the limits of integration in the equation are just the length of the defect *d* in the direction *x*.

Furthermore, the components of

$$\Delta \hat{N} = \begin{pmatrix} \Delta \hat{N}_{00} & \Delta \hat{N}_{01} \\ \Delta \hat{N}_{10} & \Delta \hat{N}_{10} \end{pmatrix} \qquad (24)$$

scale as the respective overlap integrals $O_{nn'}$ of $\delta \hat{N}_{ani}(x,z)$ with the basis functions $f_n(x)$

$$O_{nn'} = \int_0^t \delta \hat{N}_{ani}(x,z) | f_n(z) f_{n'}(z) \mathrm{d}z = \int_0^h \delta \hat{N} f_n(z) f_{n'}(z) \mathrm{d}z, \qquad (24a)$$

where *h* is the defect height in the direction *z*. Thus, we account for the 2D character of the defect.

We solve the Born equation numerically on a 1D mesh with $N_p = 100$ points placed equidistantly within the defect area 0<*x*<*d*. The Green's function of SW excitation admits an exact analytical solution in the Fourier space. The inverse Fourier transform of the analytical solution yields two components [36] – the far field in the form of a travelling SW (23) and the near field of the excitation source. No exact solution for the near field exists; but approximate solutions can be obtained for both MSSW and BVMSW cases in the form of combinations of exponents of complex arguments and exponential integrals [36,37]. Alternatively, the inverse Fourier transform can be carried out numerically on the same mesh. In this work, we employ numerical integration of the Born Equation. The process of converting the integral equation into a numerical scheme is explained in Appendix I.

The last step of this calculation is converting the amplitude of the transmitted wave into the transmission coefficient. This is done by using Eq. (2), more precisely, its modification given by Eq. (31) from Appendix I. We calculate the amplitude of the transmitted wave that enters those equations for a distance 2*d* from the rear (or far) boundary of the defect area.

More details of the extraction of the transmission coefficient from the results of the numerical calculation are given in Appendix I. Here we just briefly discuss the effect of SW losses on the value of the transmission coefficient as defined by Eq. (2) and (31). Because we normalize the amplitude of the wave transmitted through the defect area by the amplitude of the wave that travelled the same distance in a reference defect-free film, we largely eliminate the effect of SW decay on the transmission coefficient. In the areas in front of the defect and behind it, the amplitude of the forward-propagating waves just decays exponentially. The decay decrement is the same as for the wave in the reference film. This ensures that the effect of SW damping on the amplitude of the incident wave is eliminated completely. The same applies to the area past the defect. Thus, the transmission coefficient as defined by Eq. (2) accounts only for the strength of reflection of the incident wave from the defect, processes that take place within the defect area and partial conversion of the incident wave into the higher-order thickness mode. From this point of view, it does not matter where precisely we observe the transmission coefficient. Its value will be the same at any point behind the defect area, provided the point is located far enough from the far boundary of the defect, to ensure that the contribution of the near field of the far boundary to the total amplitude of magnetization-vector precession is negligible.

On the other hand, as it will be shown in Section IIC, the dispersion relation $\omega(k_x)$ for the spin waves in the defect area is different from the dispersion relation for the regular parts of the film. The spatial decay of SWs scales as $V_g^{-1}$, where $V_g = \partial\omega/\partial k_x$ is the SW group velocity. Because the group velocities are different for the defect area and the regular film, the SW decay does affect the contribution of the processes within the defect area to the transmission coefficient. What is important in this regard is the difference in the group velocities, not its absolute values. Therefore, if the difference is small, one may expect a smaller impact of the SW damping on the transmission coefficient.

In addition, below we will see that scattering of SWs from the defect is a resonant process. Before escaping the defect in the forward direction (and thus forming the transmitted wave), the wave bounces a couple of times between the defect boundaries, thus forming a partial standing wave inside the defect area. If the group velocities are similar, the effect on the transmission coefficient of the decrease in the amplitude of the wave crossing the defect area for the first time is almost fully compensated by the process of division by the denominator of Eq. (2). However, the spatial decay of the wave that is partially reflected from the far boundary of the defect area, travels back to the front boundary and then bounces back towards the far boundary is not compensated by our definition of the transmission coefficient. The transmitted wave is formed through interference of the wave that has passed the defect area one time with the wave that bounced between the boundaries and then escaped. Therefore, one expects to see a lesser amplitude of periodic oscillations of the transmission coefficient as a function of the defect "strength" for larger values of the spin wave damping coefficient or a larger length of the defect. This is because the bouncing wave will be damped more strongly and contribute less to the total amplitude of the transmitted wave. (More detail of the periodic oscillations will be given in Section III B.)

### C. Spin wave dispersion in the defect area

Performing the spatial Fourier transform (5-6) of the left-hand side of Eq. (16) and equating the resulting expression to zero yields a dispersion relation for SWs. The eigen-frequencies of SW for the defect-free (regular) areas are given by eigen-values of a matrix $D_r$ that follows from the matrix $C$ (17):

,

$$D_r = i \begin{pmatrix} 0 & -\Omega_{0k} - \omega_M P_{00} & -\omega_M Q \sin(\varphi) & 0 \\ \Omega_{0k} + \omega_M(1 - P_{00}\sin(\varphi)) & 0 & 0 & \omega_M Q \sin(\varphi) \\ \omega_M Q \sin(\varphi) & 0 & 0 & -\Omega_{1k} - \omega_M P_{11} \\ 0 & -\omega_M Q \sin(\varphi) & \Omega_{0k} + \omega_M(1 - P_{11}\sin(\varphi)) & 0 \end{pmatrix}$$

(25)

where $i$ is the imaginary unit, and the remainder of notations is explained in Appendix II. There are two pairs of the eigen-values $\pm i\omega_1(k_x)$ and $\pm i\omega_2(k_x)$, where $i$ is the imaginary unit, $\omega_1(k)$ is the frequency of the fundamental mode for a given $k_x$, and $\omega_2(k_x)$ is the eigen-frequency of the first higher-order (volume) mode for the same $k_x$.

As shown in [42], the respective matrix for a thickness-non-uniform film is obtained by adding matrix elements $\Omega_{ij}$ that represents projections of the thickness-non-uniform magnetic parameter onto the basis (14). In the case of the in-plane uniaxial anisotropy with the anisotropy easy axis parallel to the applied field, the dispersion matrix for the defect area $D_d$ reads

$$D_d = i \begin{pmatrix} 0 & -\Omega_{0k} - \omega_M P_{00} - \Omega_{00} & -\omega_M Q_{01} \sin(\varphi) & -\Omega_{01} \\ \Omega_{0k} + \omega_M(1 - P_{00}\sin(\varphi)) + \Omega_{00} & 0 & \Omega_{01} & \omega_M Q_{01} \sin(\varphi) \\ \omega_M Q_{01} \sin(\varphi) & -\Omega_{01} & 0 & -\Omega_{1k} - \omega_M P_{11} - \Omega_{11} \\ \Omega_{01} & -\omega_M Q_{01} \sin(\varphi) & \Omega_{0k} + \omega_M(1 - P_{11}\sin(\varphi)) + \Omega_{11} & 0 \end{pmatrix},$$

(26)

where $\Omega_{ij} = \gamma \, \Delta N_{ani}^{\parallel} M_s \int_0^h f_i(z) f_j(z) \mathrm{d}z$, $h$ is the defect height, $M_s$ is the saturation magnetization for the film, $\Delta N_{ani}^{\parallel}$ is the component of the tensor of the effective demagnetizing factors of anisotropy that is responsible for the extra effective anisotropy field in the defect (the effective anisotropy field is $\Delta N_{ani}^{\parallel} M_s$), and the functions $f_i(z)$ are given by Eqs. (13) and (14). The two positive eigen-values of $D_d(k)$ represent the frequencies of the fundamental and the first higher-order volume mode for the defect area.

## III. Results and discussions

### A. Comparison of MSSW and BVMSW

We first compare the MSSWs to BVMSWs. Figure 1b-c shows the magnon band structures of the system when the magnetization vector $\vec{M}$ is either perpendicular or parallel to the wave vector $\vec{k}$, respectively. The color plots are the results of micromagnetic simulations, while the dotted lines are based on analytical calculations from Ref. [43]. The case of $\vec{M} \perp \vec{k}$, named MSSWs, also known as the "Damon-Eshbach" geometry, is presented in Fig. 1b. This mode features a higher group velocity, nonreciprocal (chiral) mode profile and significant robustness against backscattering from surface topographical defects [21]. On the other hand, for $\vec{M} \parallel \vec{k}$, BVMSWs, with a reciprocal mode profile, appear as displayed in Fig. 1c. In addition, the first volume mode (VM) that is quantized over the thickness of the films is shown with green color. Due to the rather small thickness of the films, the contribution of the exchange energy to the energy of the SWs is large enough to open up a gap between the fundamental mode and the VMs, leading to the absence of any *direct* mode crossing or hybridizations [21,25,37]. Note that this gap will be even larger for thinner films.

Subsequently, we study the propagation of SWs in the presence of the surface inhomogeneities introduced above in the form of areas with locally increased uniaxial anisotropy. Note that throughout our study, we consider the modes that are exemplarily shown by large white circles in Fig. 1b-c. In this case, the frequency of the MSSW is $f = 4.64$ GHz while the frequency of the BVMSW is set to $f = 3.2$ GHz. Under this condition, the wave vector for the areas outside the defect regions is set to $k_x = 25.9$ rad/µm for both modes.

### B. Parallel surface anisotropies

We first define localized uniaxial anisotropy defects with an axis parallel to the applied field in the defect region. The size of the defect is 2 μm × 1 μm × 16 nm, which means that it is long enough to cover the entire slab in the $y$-direction while its height $h$ is only 20 % of the film thickness $t$.

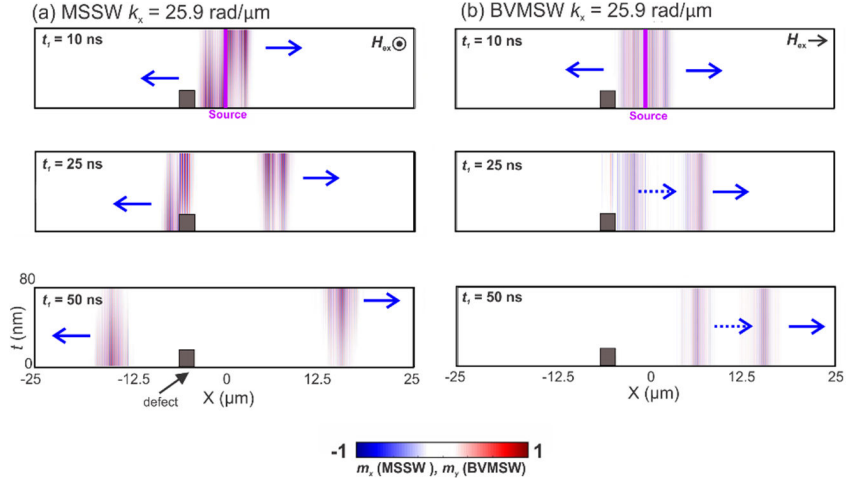

FIG 2. Snapshots of the propagating SWs in the presence of a anisotropy surface defect. The defect is shown with a grey rectangle. (a) MSSW ($f$ = 4.64 GHz and $k_x$ = 25.9 rad/μm). (b) BVMSW ($f$ = 3.2 GHz and $k_x$ = 25.9 rad/μm). Reflected and unscattered SWs are shown with dashed and solid blue arrows, respectively.

The defects are indicated with the grey rectangles in Fig. 2. They are placed on the lower surfaces of the films, where MSSWs propagating in the positive direction of the axis $x$ are localized. The anisotropy constants $K_{u1}$ are chosen such that the local effective field is enhanced within the defect - $B_{\text{eff}} = \mu_0 H_{ex} + \mu_0 H_{\text{ani}}^{\parallel} = \mu_0 H_{ex} + \frac{2K_{u1}}{M_s}$. Here, $H_{ex}$ is the external bias field, $K_{u1}$ is the first-order uniaxial anisotropy constant and $M_s$ is the saturation magnetization. This type of defect is supposed to locally shift the carrier SW number down for the carrier frequency of the incident wave packet (since we work at positions of positive group velocity for both situations, compare Fig. 1).

Figure 2 displays the snapshots of the simulated propagating SWs in the presence of the defined surface defects (grey rectangles). In this case, the defect amplitude is set to $\mu_0 H_{\text{ani}}$ = 150 mT, which is three times higher than the external field: $H_{\text{ani}}/H_{ex}$ = 3.0. As shown in Fig. 2a, the MSSWs cross the defect area almost without reflection, yielding a transmission rate close to 97%. However, under the same conditions, the BVMSWs exhibit a significant reflection when impinging on the defect, and the transmission rate drops to zero, as presented in Fig. 2b.

We next compare the MSSWs and BVMSWs more quantitatively. We fix the size of the defect as shown in Fig. 3a (top row), and plot in Fig. 3a (bottom row) the computed transmission rates obtained by micromagnetic simulations (separate data sets) for both modes as a function of the defect amplitude. Increasing the local effective field by adding a local anisotropy does not lead to a significant scattering of MSSW and the transmission rates remain above 90% even in the presence of relatively strong surface anisotropies. However, for BVMSW, the transmission rate drops below 50% if the amplitude of the defect is twice of the external field and goes to zero in the presence of a stronger effective field in the defects.

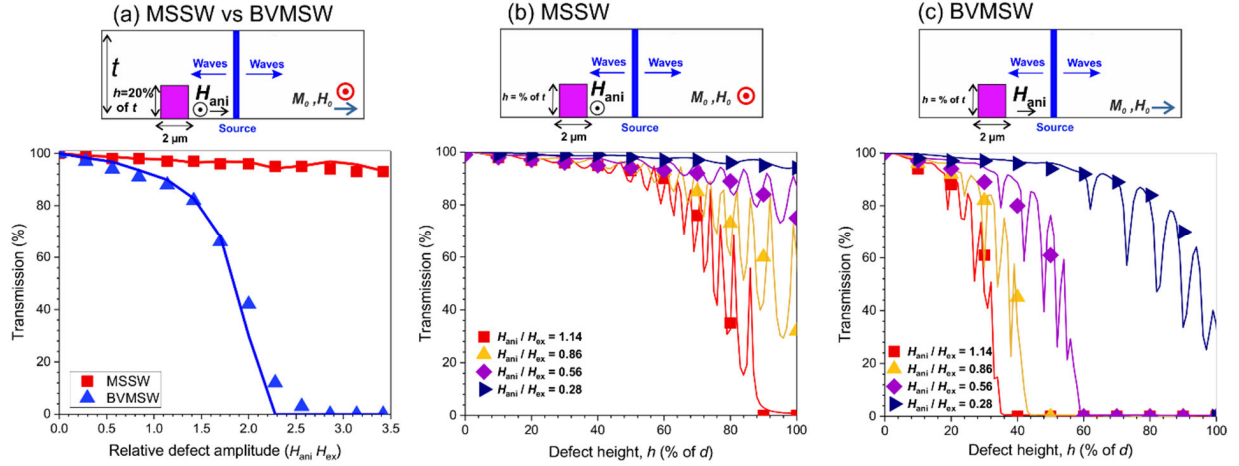

FIG. 3. Transmission of the MSSW and BVMSW as a function of the strength of local parallel surface anisotropy and corresponding sketches of geometry. Symbols indicate MuMax results, while the continuous lines are computed with our original model. (a) Transmission of MSSW and BVMSW as a function of the relative defect amplitude if the defect height is fixed to $h = 20\%$ of the film thickness $t$. (b) Transmission of MSSW as a function of the defect height for various defect amplitudes. (c) Transmission of BVMSW as a function of the defect height for a number of defect amplitudes.

Additionally, we fix the amplitudes of the anisotropy defects and vary their height $h$, as presented in Fig. 3b-c (top row). This shows how SWs are transmitted through defects that can even be as tall as the entire thickness of the film $t$. Figures 3b-c (bottom row) compare the transmission rates of MSSWs and BVMSWs with respect to the defect heights $h$ for the given defect strengths, respectively. In these figures, the discrete points are the results of the micromagnetic simulations. From these figures, we conclude the following: (*i*) for the investigated parameters, the height of the defects $h$ must be more than 50% of the thickness $t$, in order to observe a strong reflection of the MSSW, regardless of the strength of the surface defects; (*ii*) lower transmission rates of BVMSW in comparison to MSSWs in the presence of an identical defect are observed.

We use the micromagnetic results to validate our original model of Eq. (22) by applying it to the same geometries. The results of the computations employing this model are shown with solid lines in Fig. 3 (below we refer to this model as "the original model"). One sees excellent agreement of the original model with the micromagnetic simulations concerning the general trend, which illustrates the validity of the theoretical assumptions. However, one also notices from Fig. 3b-c (bottom row) that the quasi-analytical theory produces dependencies that are oscillatory and non-monotonic. This is the main difference between the two methods. The presence of multiple small peaks in the results obtained with the original model is due to formation of transmission resonances for particular values of the defect height. This is consistent with the experimental and theoretical results from Ref. [36]. These resonances are not resolved in the micromagnetic simulation since in the micromagnetic simulation the propagation of a SW pulse is modeled with a relatively small duration, and hence with a relatively broad frequency spectrum, whereas the original model use the truly single-frequency SW signals.

## C. Discussion

To explain our observations, it is necessary to consider the magnon band structure of the SWs inside the defect regions and compare it to the magnon band structure of the slab without defects. We start with the fact that we are dealing with a very thin YIG film, so let us have a closer look at the SW dispersion as shown in Fig. 1. For this range of film thicknesses, the contribution of the exchange energy to the SW dispersion relation is large and leads to two important features in the spectrum. The first one is a strong increase in the frequency of the fundamental modes (MSSWs and BVMSWs) for large wave numbers $k \gg 1/t$ where the dipolar contributions to the frequency are saturated. For the BVMSW case, this effect even leads to a change in the slope of the dispersion relation from negative for small wave numbers (where the dipole contribution to the magnon energy dominates) to a positive one at larger frequencies (where the exchange interaction provides a significant contribution to the total magnon energy). This behavior is clearly seen from Fig. 1b. The second feature that originates from the exchange-energy contribution is a strong upshift of frequencies of the higher-order thickness modes, also known as Volume Modes (VM). Those modes, characterized by the non-uniform distribution of the dynamic magnetization across the film thickness (Eq. 12), acquire large frequencies. In Fig. 1, the first volume mode is shown by green lines.

An important consequence of the exchange upshift of the higher-order-mode frequencies is that the fundamental SW mode (shown with the blue lines in Fig. 4, similar to Fig. 1) is now characterized by a much more uniform distribution of dynamic magnetization across the film thickness than for the exchange-free Damon-Eshbach type waves [44] typical for much larger film thicknesses [25,45]. As a result, the properties of the SWs propagating in the film become dependent mostly on the averages of the film parameters over the film thickness, and much less on the non-uniformity of those parameters across the film thickness [45].

In the following, we use this fact to analyze our simulation results. We consider a film with a thickness of $t = 80$ nm and assume two different values to the magnitude of the effective field of uniaxial parallel anisotropy - $H_{ani}/H_{ex} = 0.56$ and $H_{ani}/H_{ex} = 1.14$. Material parameters that are identical to the ones discussed in the Methods section are used in these calculations. Figure 4a-b show the corresponding band structures for the MSSWs and BVMSWs with and without uniaxial anisotropies, respectively. The figures represent the dispersion relations of the slab (blue curve) and inside the defect regions (black and red curves). To calculate the local dispersion for the defect, the anisotropy field is averaged over the film thickness (named as the "mean field"), and the dispersion curve is calculated for a film possessing a spatially-uniform anisotropy field equal to the mean field.

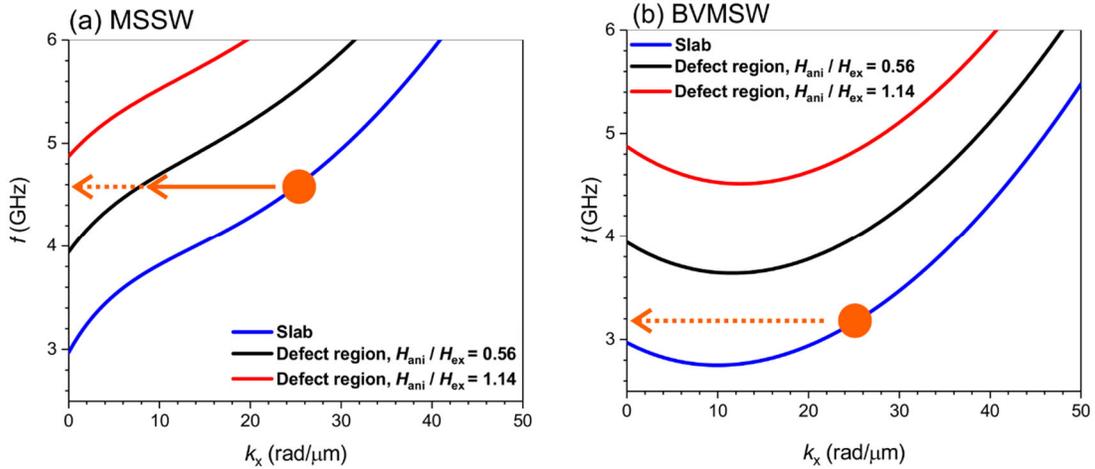

FIG. 4. Magnon band structures of the slab without defect in comparison to the band structure inside the defect regions with different anisotropies. (a) MSSWs and, (b) BVMSWs.

The carrier waves of the wave packet are shown with the orange dots in Fig 4a-b. From these panels, one sees that two regimes of the incident wave packet interaction with the defect are possible. The first one is depicted with the solid arrow in Fig. 4a. This regime corresponds to a defect which has a magnitude of $H_{ani}/H_{ex} = 0.56$. The local SW dispersion law for the defect is shown with the black line in Fig. 4a. One sees that this dispersion law is characterized by a non-vanishing solution for the wave vector at the frequency of the incident wave. (The tip of the solid arrow points to this spectral point.) The presence of the non-vanishing local wave vector evidences that propagation of SWs in the defect is allowed. Thus, while interacting with the defect, the wave scatters to a state which has a smaller momentum inside the defect, as displayed by the continued orange arrow. This two-magnon scattering process is characterized by partial reflection of the incident wave, however, partial and relatively efficient transmission through the defect is also allowed through the process of resonant scattering [36]. Indeed, in the case of the solid arrow in Fig. 4a, the transmission rate is ~76%, as shown in Fig. 3b.

The second possible regime is characterized by the local dispersion law shown with the red line in Fig. 4a. It corresponds to $H_{ani}/H_{ex} = 1.14$. The dashed arrow indicates that for this dispersion curve, there is no momentum state that corresponds to the frequency of the incident wave. As a result, the only way for the packet to cross the defect is through the process of tunneling [36,46]. If the length of the defect is large enough, the efficiency of the tunneling process is negligible, and this mode is fully reflected from the front edge of the defect. The same reasoning applies to the BVMSW case – however, here a full reflection must already take place in the presence of a defect with an amplitude equal to $H_{ani}/H_{ex} = 0.56$ (purple curve in Fig. 3c, dashed arrow in Fig. 4b), because no resonant scattering to a mode inside the defect is possible.

Now recall our claim that the scattering properties of a defect are mostly determined by the defect magnitude averaged over the film thickness. From Fig. 3b one sees, that a MSSW wave packet becomes fully reflected from the defect with a magnitude $H_{ani}/H_{ex} = 1.14$ when the defect height reaches 90% of the total film thickness. This implies that the average anisotropy field $H_{ani}/H_{ex} = 1.14 \times 0.9 = 1.026$. Our calculations show that the MSSW dispersion relation for a thickness-uniform $H_{ani}/H_{ex} = 1.026$ has a cut-off frequency $f(k_x=0) = 4.69$ GHz that is very close but below the frequency of the incident wave packet. This is in agreement with the vanishing transmission for $H_{ani}/H_{ex} > 1.026$ from Fig. 4a.

The situation is slightly more complicated in the case of BVMSW (Fig. 4b) – the minimum frequency for the SW band does not correspond to a mode at $k_x = 0$. The minimum now sits at $k_x \sim 10$ rad/μm. Despite this difference, we see the same trend for the BVMSW configuration – the defect area stops transmitting the SW signal for the defect height of 40% that corresponds to the value of mean field $H_{ani}/H_{ex} = 1.14 \times 0.4 = 0.46$. Its dispersion gives $f(k_x = 10$ rad/μm$) = 3.48$ GHz. This value is very close but slightly below the frequency of the incident wave packet, but it is enough to block the SW transmission.

From Fig. 4 one also clearly sees that the frequency of the incident wave packet is noticeably closer to the bottom of the BVMSW spectrum than to the MSSW one. This implies that a smaller defect height is able to push the packet carrier frequency locally out of the BVMSW SW band. As a result, a relative defect height of 0.46 is enough to fully block the BVMSW propagation through the defect area, but a much taller defect (of 90%) is needed to stop transmission of MSSW completely.

To confirm this idea, we plot the efficiency of SW transmission through the defect satisfying $H_{ani}/H_{ex} = 1.14$ and the upshift of the SW dispersion law in the defect in Fig 5. The results of micromagnetic simulations and the data obtained with the original theory are shown as separate red data sets and continuous red lines, respectively. The blue dashed and dotted lines show the dispersion shift (right vertical axis). The dispersion shift is calculated as the frequency difference between the frequency of the incident wave and the frequency of the bottom of the spectrum in the defect region. The frequency for the spectrum's bottom is obtained by solving the eigen-value problem for the matrix $D_d(k)$ (26) that is a slightly more accurate approach than using the average defect field.

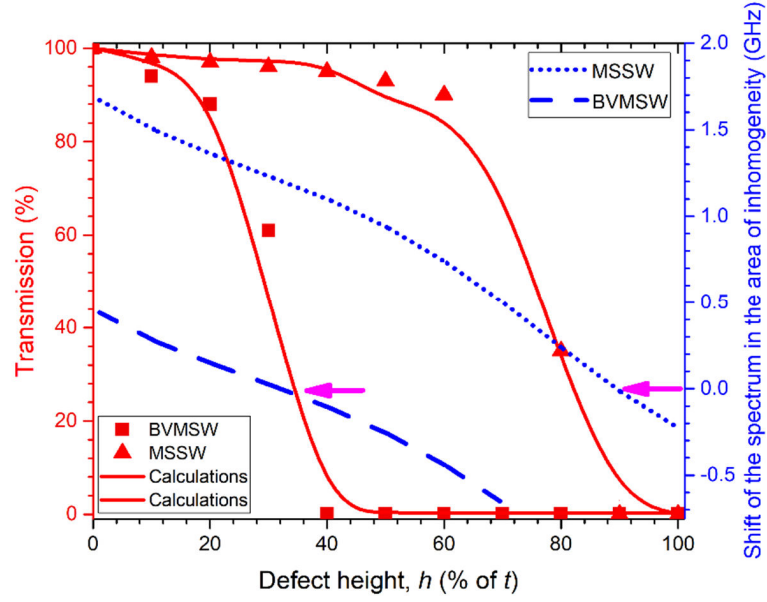

FIG. 5. Left axis (red color coded): spin-wave transmission of the MSSW and BVMSW as a function of the defect height. The strength of the defects satisfies $H_{ani}/H_{ex} = 1.14$. Separated data points indicate the results from MuMax micromagnetic simulations while the continuous lines are computed with our original model. Right axis (blue color coded): the shift of the spectrum of the MSSW and BVMSW inside the defect calculated with expressions (26) in comparison to the reference film (25) as a function of the defect height. The purple arrows indicated defect heights in which the energy barrier appears.

One sees a very strong correlation between the two dependencies – the SW transmission rate drops to zero in the vicinity of the defect height for which the frequency of the incident SW signal is equal to the minimum frequency (i.e. the energy gap) for the respective spectrum. This energy barrier appears when the shift of the spectrum with respect to the reference mode reaches zero, as indicated by the purple arrows in Fig. 5.

We finally note that, in the presence of several equivalent defects, the total transmission rate of the SWs may be estimated as a product of the respective transmission rates of SWs though the individual defects. Furthermore, the position of those defects with respect to the $z$ coordinate does not influence significantly the reflection and transmission coefficients. This estimation is valid if the defects are placed arbitrarily with respect to the $x$ co-ordinate. Conversely, if several defects are placed equidistantly one expects formation of a magnonic crystal of finite length. This will translate in appearance of band gaps in the frequency dependence of the transmission characteristics for the defect area similar to ones observed in [47].

So far, we have discussed the impact of defects whose uniaxial anisotropies are parallel to the external field. In reality, defects can also take the form of a change of the effective magnetization. In the following, we briefly discuss such a defect.

### D. Perpendicular surface anisotropies

As the second type of anisotropy defect, we locally introduce a perpendicular anisotropy $H_k^\perp$ in the defect region. Therefore, we define a local uniaxial anisotropy with an axis perpendicular to the applied field and normal to the film plane in the defect region, as shown in Fig. 6a. The size of the defect is 2 μm × 1 μm × 16 nm (similar to the parallel anisotropy defects). Note that we again use the same parameters as discussed before.

The resulting transmission characteristics of the SWs are shown in Fig 6a (bottom row). As before, changing the local perpendicular anisotropy does not significantly influence the transmission of MSSW, whereas the scattering via the mechanisms mentioned earlier leads to a strong drop of the transmission for BVMSW.

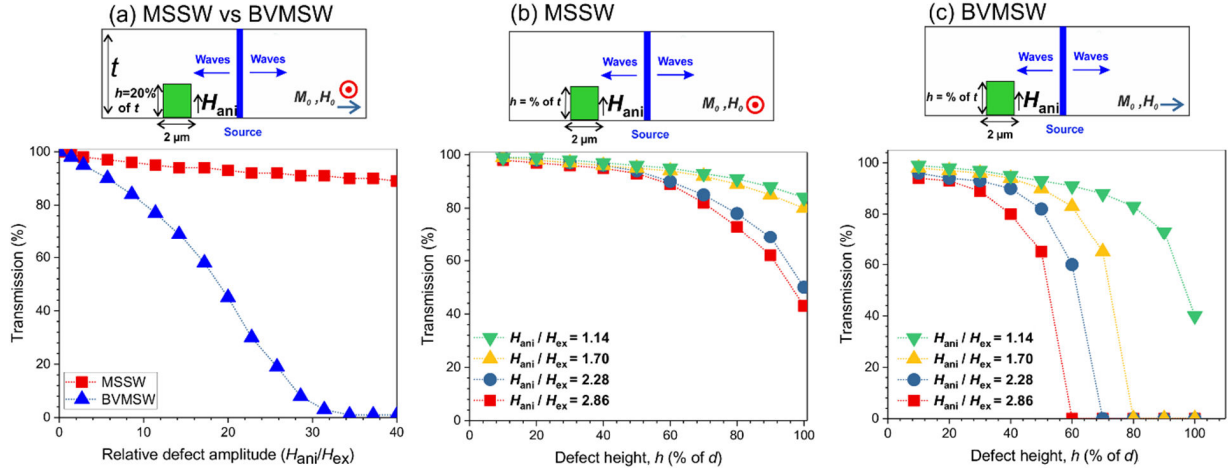

FIG 6. Spin-wave transmission of the MSSW and BVMSW as a function of the local perpendicular surface anisotropy strength (defect) and their corresponding schematic pictures. (a) Transmission of MSSW and BVMSW as a function of the relative defect amplitude if the defect height $h$ is fixed to 20% of the film thickness $t$. (b) Transmission of MSSW as a function of the defect height in the presence of different defect amplitudes. (c) Transmission of BVMSW as a function of the defect height in the presence of different defect amplitudes. Separated data points indicate the results from micromagnetic simulations and dashed lines are guides to the eyes.

To further confirm the higher robustness of the MSSW against backscattering from surface inhomogeneities in comparison to BVMSW as explained earlier, we fix the anisotropy defect amplitudes and vary their height $h$, as presented in Fig. 6b-c (top row). Figure 6b-c (bottom row) compare the transmission rates of MSSWs and BVMSWs with respect to the defect heights $h$ for the given defect strengths, respectively. Similar to the parallel anisotropy defects, it is clear that the height of the defects $h$, must be more than 50% of the thickness $t$ to ensure a strong reflection of the MSSW regardless of the strength of the surface defects. Moreover, BVMSWs in comparison to MSSWs exhibit a stronger reflection in the presence of similar defects. The underlying mechanism of these reflections is the same as the one discussed in the preceding section.

Concluding this section, we propose that the defects of the type investigated in this work may be implemented by using, for example, voltage-controlled magnetic anisotropy (VCMA). This effect allows one to tune the anisotropy of the system locally [10-12]. In addition, one may employ ion irradiation to modify the saturation magnetization for the material at the position of the defect [48-49]. Furthermore, using locally induced external magnetic fields, such as the Oersted field created by a current carrying wire on top of the thin film, is expected to produce the same effect [50].

## IV. Conclusion

We have used micromagnetic simulations and quasi-analytical calculations to present a new way to control the transmission characteristics of dipole-exchange SWs in thin films using surface inhomogeneities. Employing locally introduced uniaxial anisotropies in the form of surface defects, we have shown the criterion for controlling the SW transmission signals in real space. The small thickness of the investigated film and the high contribution of the exchange energy to the magnon band allowed us to average the anisotropy field over the film thickness, in order to describe the transmission and reflection mechanisms of the SWs. The comparison of the magnon band structure inside the defect area to the reference thin film clarifies the condition of resonant transmission via linear scattering processes, and reflection due to the energy conservation laws. Our results indicate that BVMSWs, which propagate parallel to the static magnetization vector, are more susceptible to backscattering from anisotropy defects in comparison to MSSWs that propagate perpendicular to the magnetization vector. A defect as large as the entire film thickness and the strength of which satisfies $H_{ani}/H_{ex} = 0.56$ is strong enough to block transmission of BVMSWs entirely. However, the defect strength must be twice larger and satisfy $H_{ani}/H_{ex} = 1.14$ to fully block propagation of MSSWs. We finally emphasize that the presented analysis and argumentations cannot be applied to topographical defects [21], since reducing the thickness of the film does not shift the frequency of the entire magnon band structure in a way similar to the anisotropy defects. Our study provides a new way to tune the transmission rates of SWs in thin films, and will be helpful for designing nanoscopic magnonic devices for analogue and digital data processing, where SWs are used as data carriers.

**Acknowledgment**

This project was funded by the Deutsche Forschungsgemeinschaft (DFG, German Research Foundation) – TRR 173 – 268565370 (project B01), by the DFG through the DU 1427/2-1 project, by the Nachwuchsring of the TU Kaiserslautern and a Research Collaboration Award from the University of Western Australia. M. K. acknowledges his sabbatical leave from the University of Western Australia.

**Appendix I: Numerical Solutions of the Born equation**

The whole procedure of numerical solution of (22) is implemented as a MathCAD worksheet. It works as follows: First, the values of the Green's function $\hat{G}_{exc}(x_i)$ are calculated for the mesh points $i=0,1 \ldots N_p$ by performing numerical inverse Fourier transformation of the analytical expression for its Fourier image. Second, values of the tensor product $\hat{G}_{exc}(x_i - x_j)\Delta\hat{N}\,\Delta x$ are computed for the mesh points. (Here $\Delta x$ is the mesh cell size.) This converts the tensor into a matrix $\hat{\Theta}$ with elements $\Theta_{ij}$. Then the right-hand side of (22) is obtained by acting with $\hat{G}_{exc}(x)$ on an assumed localized excitation source. The source is located at a single point $x_0$ in front of the defect. It is separated from the defect by a distance that is large enough to ensure that the front edge of the defect is situated in the far zone of the source. The source has only one vector component of the driving field. It is the $z$ one. This choice is dictated by the fact that this component can drive both BVMSW and MSSW. Thus, the external source is

$$|\mathbf{h}_{dr}(x_0)\rangle = \begin{pmatrix} 0 \\ h(x_0) \end{pmatrix}$$

As a result, the discrete version of the right-hand side of Eq. (22) takes the form of a simple product $\hat{G}(x_i - x_o)|\mathbf{h}_{dr}(x_0)\rangle$ that represents a column vector $\overline{\Gamma}$ with components $\Gamma_i$.

These three steps allow us to convert the integral equation into an inhomogeneous vector-matrix equation

$$\hat{\Theta}\,\overline{M} = \overline{\Gamma}, \qquad (27)$$

where $\overline{M}$ is an unknown column vector having $|\mathbf{m}(x_i)\rangle$ as its components.

We solve the matrix equation (27) using the numerical methods of linear algebra built in the MathCAD software. The solution represents values of $|\mathbf{m}(x)\rangle$ at the points $x_i$ within the defect area. In order to calculate the transmission coefficient, we pick up an "observation point" $x_t$ located behind the defect and at a distance $2d$ from its rear boundary. At this distance, we expect only traveling waves with characteristics corresponding to eigen-waves of the regular (homogeneous) film to be present. Put differently, we expect all potential near-field effects such as leaky waves localized at the far boundary of the defect not to reach $x_t$ because of its significant distance from the defect. The SW amplitude at the observation point is computed using the discrete version of the same equation (22)

$$|\mathbf{m}(x_t)\rangle = -\Delta x \sum_{i=1}^{N_p} \hat{G}_{exc}(x_t - x_i)\Delta\hat{N}|\mathbf{m}(x_i)\rangle + |\mathbf{m}^0(x_t)\rangle \exp(-ik_x x_t)\exp(-\nu x_t), \qquad (28)$$

where, as before, $x_i$ are the mesh points within the defect area.

We assume that the incident wave $|\mathbf{m}^0(x)\rangle$ represents a fundamental mode of SWs of the film. For symmetry reasons, if the defect height $h$ is smaller than the film thickness $t$, we may expect scattering of the fundamental mode into the first higher-order (anti-symmetric) SW mode. Therefore, in the general case, the dynamic magnetization at the observation point $|\mathbf{m}(x_t)\rangle$ represents a combination of the fundamental and the first higher-order mode. We separate the contributions based on the fact that the eigen-waves of the regular film are characterized by ratios $|\mathbf{m}_0(x)/\mathbf{m}_1(x)|$ (see Eq.(12)) that are specific to particular eigen-modes. This implies that the total SW "field" $|\mathbf{m}(x_t)\rangle$ is

$$|\mathbf{m}(x_t)\rangle = A_f \begin{pmatrix} \mathbf{m}_0^f \\ \mathbf{m}_1^f \end{pmatrix} + A_{1ho} \begin{pmatrix} \mathbf{m}_0^{1ho} \\ \mathbf{m}_1^{1ho} \end{pmatrix}, \qquad (29)$$

where $A_f$ and $A_{1ho}$ are the sought scalar amplitudes of the fundamental and the 1st exchange mode respectively, and the vectors

$$\begin{pmatrix} \mathbf{m}_0^f \\ \mathbf{m}_1^f \end{pmatrix} \quad \text{and} \quad \begin{pmatrix} \mathbf{m}_0^{1ho} \\ \mathbf{m}_1^{1ho} \end{pmatrix} \qquad (30)$$

are the vectors typical for the respective eigen-waves for the film. For instance, if the film is thin to the extent that there is no signature of anti-crossing (hybridization) of the two modes, $\mathbf{m}_1^f = 0$, $\mathbf{m}_0^f = (m_x^f, m_z^f) = (1, m_z^f)$ for the fundamental eigen-wave, and $\mathbf{m}_0^{1ho} = 0$ and $\mathbf{m}_1^{1ho} = (1, m_z^{1ho})$ for the eigen-waves representing the first exchange mode. If a signature of mode hybridization is present in the eigen-wave spectrum, the situation is technically more complicated, but remains qualitatively the same.

The vectors (30) are computed as eigen-vectors of the direct spatial Fourier transform of matrix $C$ (27) and then properly normalized to obtain $m_x^{f(1ho)} = m_x^{1ho} = 1$. Equation (29) represents a system of linear equations from which the unknown amplitudes $A_f$ and $A_{1ho}$ are found. The transmission coefficient $T$ is then defined as

$$T(h) = A_f(h) / A_f(h=0). \quad (31)$$

Recall that $h$ is the defect height, and hence, $h=0$ corresponds to a film without any defect present.

We repeat that the model is implemented as a MathCAD worksheet. It takes about 8 minutes to complete calculation of $T$ for 100 values of $h$ covering the range from zero to $h=t$.

**Appendix II: Quantities that enter equations (25) and (26).**

$$\Omega_{nk} = \omega_H + \omega_M \alpha k_n^2, \quad P_{00} = 1 - \frac{1-\exp(-|k_x|t)}{|k_x|t}, \quad P_{11} = \frac{k_1^2 - \frac{2|k_x|}{t}[1-\exp(-|k_x|t)]}{2k_1^4},$$

$$Q = \frac{\sqrt{2}k_x^2}{k_1^2}[1-\exp(-|k_x|t)], \quad k_n^2 = k_x^2 + \frac{n^2\pi^2}{t^2}, \quad n=0,1, \quad \omega_H = \gamma H, \quad H \text{ is the applied field}, \quad \omega_M = \gamma M_s, \quad M_s \text{ is}$$

the saturation magnetization for the film, $\alpha$ is the exchange constant, and $\varphi=0$ for BVMSW and 90 degree for MSSW.

**Correspondence**: *mohseni@rhrk.uni-kl.de